\documentstyle[epsfig,12pt]{article}
\begin{document}
\begin{titlepage}
\hfill INP MSU 98-13/514

\hfill March 23, 1998

\vspace*{3cm}
\begin{center}
\Large \bf Higgs Boson Discovery Potential of LHC \\ 
in the Channel $pp \to\gamma\gamma+jet$
\end{center}

\vskip 2cm

\begin{center}
\large
        S.~Abdullin$^+$,
        M.~Dubinin$^*$, 
        V.~Ilyin$^*$, \\
        D.~Kovalenko$^*$,
        V.~Savrin$^*$,
        N.~Stepanov$^\sharp$
\end{center}

\begin{center}
 { \it $^+$ UHA, Mulhouse, France (on leave of absence ITEP, Moscow, Russia)} \\
 { \it $^*$ Institute of Nuclear Physics, Moscow State
                  University, 119899 Moscow, Russia} \\
 { \it $^\sharp$ Institute for Theoretical and Experimental Physics,
                  117259 Moscow, Russia}   
\end{center}

\begin{quotation}

\vskip 1.5cm


\vskip 2cm
\begin{center}
{\sf Abstract}
\end{center}

\vskip 0.5cm

We discuss the SM Higgs discovery potential of LHC in the reaction $pp\to H+
jet\to\gamma\gamma+jet$ when the jet is observed  at sufficiently high $E_t$ to
be reliably identified. We conclude that this channel gives promising discovery
possibilities  for the Higgs boson mass range 100-140 GeV, during LHC operation
at a low  luminosity. With 30 fb$^{-1}$ of accumulated data and for $M_H=120$
GeV about 100 signal events could be observed  with the  number of background
events larger by a factor of 2 only, showing a signal  significance
$S/\sqrt{B}\sim 7$. We use the difference of distributions in the partonic
subprocess energy  $\sqrt{\hat s}$ for the signal and background for a better
separation of the signal.

\end{quotation}

\end{titlepage}
\noindent
{\large \bf Introduction}
\vskip 0.2cm

\noindent 
It is well known that the observation of Higgs boson with mass
$M_H<140$ GeV at the LHC collider (pp,\,$\sqrt{s}=$14 TeV) in the inclusive
channel $pp\to\gamma\gamma+X$ is not easy \cite{atlasTDR,cmsTDR}. The
$\gamma\gamma$ continuum rapidly increases for smaller $\gamma\gamma$ pair
invariant masses, and it is necessary to separate a rather elusive Higgs boson
signal from it. In this situation it is important to understand whether we can
observe any other discovery channels. In this paper we are considering (in the
Standard Model) the reaction $pp\to H+jet\to\gamma\gamma+jet$ when Higgs
boson is produced  with large transverse momentum recoiling against a hard jet.
Of course, in this channel the signal rate is much smaller in comparison with
the inclusive $pp\to\gamma\gamma+X$ case. At the same time, as we shall see
below, the situation with the background is undoubtedly much better.  We can
usefully exploit richer kinematical features of the final state
$\gamma\gamma+jet$, when some specific jet distributions in the partonic c.m.s.
are different for signal and background processes.

The idea to look for Higgs boson associated with a high $E_t$ jet in the
final state was considered in \cite{j1}. In \cite{Kao,alan} the corresponding
subprocesses were calculated within the minimal supersymmetric extension of the
Standard Model. However, in \cite{Kao,alan} Higgs decay channels were not
considered and in \cite{j1} only the final state $\tau^+\tau^- +jet$ was
analyzed. In \cite{j2} the SM heavy Higgs boson decay into the $WW$ or $ZZ$
pairs was discussed for $H+jet$ production. Promising results have been
obtained recently in \cite{RainZepp} for $\gamma\gamma+2jets$ final state with
two very forward jets ($|\eta|<5$). In \cite{abdullin} the final state
$\gamma\gamma+(\geq 2jets)$ was simulated in the realistic CMS detector
environment. The final state $\gamma\gamma+jet$ with only one high $E_t$ jet
recoiling against the Higgs boson has not been analyzed yet and we investigate
it in detail \footnote{Primary analysis of the reaction $pp\to\gamma\gamma+jet$
can be found in \cite{CMSournote}, where we also considered the processes
$pp\to WH,t\bar tH \to \gamma\gamma+lepton$.}.

\vskip 0.5cm
\noindent
{\large \bf 1. Calculation framework}
\vskip 0.2cm

\noindent
We calculated cross sections and distributions by means of CompHEP package
\cite{comphep}. Methods of multichannel adaptive Monte Carlo integration over
the phase space implemented in the package are described in \cite{IKP}.

The cross sections of signal subprocesses under discussion depend significantly
on the choice of QCD parameters.  There are three sources of this
dependence: (1) QCD evolution of the parton distribution functions; (2)
$\alpha_s$ dependence in the subprocesses;  (3) $H\to\gamma\gamma$ branching 
variation due to QCD corrections to the $\Gamma^{tot}_H$. In the leading order
the corresponding corrections can be factorized. Moreover, due to a very small
value of $\Gamma^{tot}_H$ for $M_H<140$ GeV the fixed value of strong coupling
$\alpha_s(M_H)$ can be used for the evaluation of $H\to\gamma\gamma$ branching.
However, it is well-known that for the reaction $pp\to H\to\gamma\gamma$ the
dependence on the renormalization scale $\mu$ and on the parton factorization
scale $Q$ is strong enough, and the next-to-leading order analysis is necessary
(see \cite{SDGZ} and references therein). The NLO corrections decrease this
$(\mu,Q)$ theoretical uncertainty, showing only a $\sim 15$\% sensitivity  of
the final result. One can hope to observe a similar effect also for the case of
Higgs boson production at high $p_t$. However, self-consistent analysis
requires the NLO corrections to hard subprocesses which are not known yet,
unfortunately\footnote{Recently formulas for the corresponding virtual
corrections were calculated in the $m_t\to \infty$ limit \cite{Carl}.}. Thus,
today we made only the LO analysis when parton distribution functions and 
running $\alpha_s$ in subprocesses are taken at the leading order. We used the
parametrization CTEQ4l ($\alpha_s(M_Z)=0.1317$, $\Lambda_{QCD}^{(5)}=181$ MeV)
\cite{cteq4}. At the same time QCD NLO formulas \cite{MQrunning,SDGZ} were
used for the evaluation of $\Gamma^{tot}_H$ with the reference value
$Br(H\to\gamma\gamma)=1.534\cdot 10^{-3}$ at $M_H=100$ GeV. Parameter values
used in our analysis are $M_Z=91.187$ GeV, $\sin{\theta_w}=0.4732$, and
$m_s=0.2$ GeV, $m_c=1.42$ GeV, $m_b=4.62$ GeV, $m_t = 175$ GeV.

Typical acceptances of the LHC detectors ATLAS and CMS should be taken into
account in the analysis. For photons we are using the cuts similar to already
well analysed cuts of the inclusive channel \cite{atlasTDR,cmsTDR}:  two
photons are required with $p_t^\gamma>40$ GeV  for each photon (somewhat harder
than for the inclusive channel), and photon rapidity $|\eta_\gamma|<2.5$. 
For a jet two sets of basic kinematical cuts will be discussed in further 
analysis:

\vskip 1mm
\centerline{
{\bf C1:} $E_t^{jet}>40$ GeV,  $|\eta_{jet}|<2.4$;
\hspace{2.5cm} {\bf C2:}  $E_t^{jet}>30$ GeV,   $|\eta_{jet}|<4.5$.
}
\vskip 1mm

\noindent
The first set corresponds to the central part of calorimeter where a jet with
transverse energy greater than 40 GeV can be detected with  highest efficiency.
The second set assumes forward parts of the hadron calorimeter involved.  The
analysis of background processes made for the inclusive channel has shown that
photons should be isolated. So we apply the cut $\Delta R>0.3$ for each
$\gamma-\gamma$ and  $\gamma-g(q)$ pair, where $\Delta
R=\sqrt{\Delta\phi^2+\Delta\eta^2}$ is the standard variable separating 
particles in the {\it azimuth angle -- rapidity} plane.

\vskip 0.5cm
\noindent
{\large \bf 2. Signal processes}
\vskip 0.2cm

\noindent

There are three QCD subprocesses giving a signal from the Higgs boson in the
channel under discussion: $gg\to H+g$, $gq\to H+q$ and $q\bar q\to H+g$.
Feynman diagrams contributing in the leading order $\alpha_s^3$ are shown in
Fig.~\ref{fig:fd_s_QCD}. The corresponding matrix elements were calculated
analytically in \cite{j1} and we implemented these formulas in our code. We
found that the $gg\to H+g\to\gamma\gamma+g$ subprocess gives the main
contribution to the signal rate. The $gq$ channels with light quarks give
20-25\% of the {\it gluon-gluon} signal contribution, while the $q\bar q$
channels can be neglected. Furthermore, at large values of parton factorization
scale, $\sim 10^4$ GeV$^2$, typical for intermediate Higgs boson production,
one can expect noticeable contributions of the strange and heavy quark sea. We
found that it gives 10\% of the dominant {\it gluon-gluon} subprocess. Then,
tree-level subprocess $gg\to H+b+\bar b$, where the Higgs boson is radiated
from the final $b$ quark or $b$-propagator, contributes to $\gamma\gamma+ jet$
signal events if we do not register ({\it veto}) the events with both $b$ quark
jets observed. We found that this process gives about 1.5\% of the $gg$
contribution. Finally we note that the s-channel subprocesses 
$q\bar q\to H+Q+\bar Q$, where $q$ is a light quark and $Q$ is $b$ or $t$ 
quark, are negligible. In total QCD subprocesses give 3.3 fb, 5.7 fb and 
5.5 fb with the {\bf C1} set of cuts for $M_H=100$, 120 and 140 GeV, 
correspondingly, and  5.5 fb, 10.6 fb and 9.8 fb  with the {\bf C2} set.

The second group of signal subprocesses includes the electroweak reactions of
Higgs production through $WW$ or $ZZ$ fusion (Fig.~\ref{fig:fd_s_WZ}a) with the
high $E_t$ jet in the final state from the scattered quark, and the Higgs boson
production associated with $W$ or $Z$ (Fig.~\ref{fig:fd_s_WZ}b) decaying into
quark-antiquark pairs. Of course, if the signature with only one jet is under
discussion, one should {\it veto} the second quark jet. The $WW/ZZ$ fusion
processes were calculated in \cite{Daw84-KRR84}, while the processes of $HW/HZ$
associated production were considered in \cite{e1}. One should note that the
contribution of these subprocesses decreases when we change {\bf C1} set of
cuts to {\bf C2}, what is opposite to the case of QCD signal subprocesses. The
reason is {\it veto} condition for one of the jets. Indeed, typical transverse
momentum of the final quark (in the dominant $WW/ZZ$ fusion production
processes) is about half of the vector boson mass. So, weaker jet cut $E_t>30$
GeV in the {\it veto} condition removes a larger number of events. We found
that the cross section of fusion processes  is 2-3 times larger than for the
processes of associated production. In EW channels $s$ and $c$ quark sea gives
about 10\% of $u$ and $d$ quark (valence plus sea) contribution. In total, EW
processes give 1.0 fb, 1.6 fb and 1.5 fb with the {\bf C1} set of cuts for
$M_H=100$, 120 and 140 GeV, correspondingly, and 0.65 fb, 1.1 fb and 1.1 fb
with the {\bf C2} set. So the electroweak signal rate is on the level of 30\%
of the QCD signal  with the {\bf C1} set of cuts and 10\% with the {\bf C2} set.

\vskip 0.5cm
\noindent
{\large \bf 3. Irreducible background processes}
\vskip 0.2cm

\noindent 
Let us now look at the irreducible background.  Three subprocesses 
contribute here, two of them at the tree level, $gq \to \gamma+\gamma+ q$ 
(Fig.~\ref{fig:fd_Birr}a) and $q\bar q\to\gamma+\gamma+g$
(Fig.~\ref{fig:fd_Birr}b), and the third one at the one loop level,
$gg\to\gamma+\gamma+g$. The $q\bar q$ channel with $u$ and $d$ quarks in the 
initial state gives the cross section on the level 15-20\% of the main
irreducible background $gq$  ($q=u,d$) with the {\bf C1} set of cuts, and 
15-30\% with the {\bf C2} set for Higgs mass range 100-140 GeV. The
contributions of strange and heavy quark sea in the initial state are on the
level 25-30\% of the main $gq$ channel. Here about 85\% of the contribution is
coming from the $c$ quark sea. The $s$ quark contribution is suppressed by the
factor 16 originating from a smaller fractional charge.

The matrix element for the one-loop process $gg\to\gamma+\gamma+g$, when the
photons are  radiated from the quark loop, is unknown yet. We estimated this
cross section by means of PYTHIA \cite{pythia} simulation, switching on the
gluon bremsstrahlung from the initial state in the subprocess
$gg\to\gamma+\gamma$. This is definitely only one of physical contributions to
the one-loop process. The result of these simulations shows us that the
one-loop background is about 2-4\% of the main contribution coming from the
$gq$ channel. So we neglect the one-loop background in further analysis,
however, this point is one of a serious theoretical uncertainties.

In total, the irreducible background contribution amounts to 9.3 fb, 13.7 fb
and 16.1 fb in the 1 GeV bin around the values of $M_{\gamma\gamma}=$ 100, 120
and 140 GeV, correspondingly, and with the {\bf C1} set of cuts.  Of course, 
with the {\bf C2} set of cuts the background is larger,  the corresponding
numbers are 15.3 fb, 25.3 fb and 25.1 fb in the 1 GeV bin.

\vskip 0.5cm
\noindent
{\large \bf 4. Reducible background processes}
\vskip 0.2cm

\noindent
Various processes could give a background due to radiation  of photons from the
fragmentating quarks or gluons. Photon production in a jet hadronization is
also possible, and is defined in particular by a $\pi^0$-meson production. The
energetic $\pi^0$'s, decaying mainly to a photon pair, can be detected as a
single photon in the electromagnetic calorimeter. First kind of such reducible
background is coming from the subprocesses  $gq\to\gamma+g+q$,
$gg\to\gamma+q+\bar q$ and $qq'\to \gamma+q(g)+q'(g)$, in the case when the
final gluon or quark produces an energetic isolated photon but other products
of hadronization escape the detection as a jet. One can say that {\it a jet is
misidentified as a photon}. Second kind of reducible background could come from
the subprocesses $gq\to\gamma+q$, $q\bar q\to\gamma+g$ when the second photon
is produced in the quark or gluon fragmentation but other products of the
hadronization are still detected as a jet with proper separation from this
photon. Third source of reducible background is connected with the pure QCD
subprocesses of $2\to 2$ and $2\to 3$ types, when all particles in the final
state are gluons or quarks. We performed rough analysis of these QCD reducible
backgrounds in \cite{CMSournote} and found that they turn out to be less than
20\% of the irreducible background, and the misidentification rate is given
mainly by the processes of the first kind. Here we used the $\gamma(\pi^0)/jet$
rejection factor equal to 2500 for a jet misidentified as a photon and 5000 for
a well separated $\gamma(\pi^0)$ production by a jet. These factors were
obtained for a jet satisfying the cuts described above with the help of PYTHIA
simulations. No additional $\pi^0$ rejection algorithms were used.

\vskip 0.5cm
\noindent
{\large \bf 5. Dependence on the $Q^2$ scale}
\vskip 0.2cm

\noindent
In our calculations we used $Q^2=M_H^2+2(E_t^{jet})^2$ as the parton
factorization scale and the normalization scale for running $\alpha_s$ in the
hard QCD signal subprocesses, as well as in the background processes. At the
same time for the $WW/ZZ$ fusion we used $Q^2=(M_V/2)^2$ ($M_V=M_W,M_Z$) and
$Q^2=(M_V+M_H)^2$ for the $HW/HZ$ associated production. Certainly, due to
$\alpha_s^3$ order of partonic subprocesses  one can expect strong dependence
on the choice of $Q^2$ for the QCD signal. We checked that for $Q^2=M_H^2$, the
corresponding signal cross section increases by 15\%, while the background
cross section increases only by $\sim 5$\%. If one uses $Q^2=(50\mbox{GeV})^2$,
the QCD signal cross section increases by 80\% while the background only by
13\%. Such strong $Q^2$ dependence shows that the complete NLO analysis is
needed.

\vskip 0.5cm
\noindent
{\large \bf 6. Reconstruction of event kinematics}
\vskip 0.2cm

\noindent 
The distributions $d\sigma/d\sqrt{\hat s}$ in parton collision energy presented
in Fig.~\ref{fig:aaj-shat} show that the background processes contribute at a
smaller $\sqrt{\hat s}$ in comparison with the QCD signal processes. So, one
can hope that the corresponding cut can improve the {\it S/B} ratio. Photon
energy can be measured with a high enough presision, so the main uncertainty of
the $\sqrt{\hat s}$ reconstruction is defined by the  accuracy of a final
parton energy and momentum reconstruction (see details in
\cite{atlasTDR,cmsTDR}). The energy of a parton can be reconstructed with the
accuracy $(\delta E/E)^2\sim 100\%\sqrt{E}\oplus5\%$. It means that for jets
with transverse energy more than 30 GeV the parton energy will be reconstructed
with the accuracy $\sim 7$ GeV. Taking into account all factors, one can expect
that $\sqrt{\hat s}$ variable can be reconstructed with the error $\sim 10$
GeV. It is clear from the distributions presented in Fig.~\ref{fig:aaj-shat}
that such accuracy is good enough to apply a cut on $\sqrt{\hat s}$ for the
suppression of the background. However $\delta E$ is mostly related to the
energy loss due to the undetected products of hadron fragmentation. It follows
that the experimental distributions can be shifted to the smaller $\sqrt{\hat
s}$ values in comparison with Fig.~\ref{fig:aaj-shat}. But the scale of
smearing in this distribution (more dangerous for our analysis) should be
definitely smaller than 10 GeV. One should note that processes at higher order
$\alpha_s$ (contributing to the NLO corrections) have more than 3 particles in
the  partonic final state. So, strictly speaking, the energies and momenta of 
photons and the detected jet are not sufficient for the reconstruction of 
partonic c.m.s. collision energy in these events. However, one can hope that
the main contribution to the QCD corrections should go from virtual and soft
gluons and cannot affect significantly the reconstruction of $\sqrt{\hat s}$
variable.

We found that the cut $\sqrt{\hat s}>210$ GeV decreases the QCD signal cross
section only by 25-30\% with the {\bf C1} set of basic cuts, while the
background is suppressed much stronger, for example by a factor of 2 when
$M_H=120$ GeV. For the {\bf C2} set of basic cuts the corresponding numbers
are 20-35\%  decrease for the signal rate and suppression factor of 1.7-2.8
for the background. It means, for example, that with the {\bf C2} set of cuts
the {\it S/B} ratio is improved by a factor of $\sim 2$ for $M_H=100-140$
GeV.  One can get {\it S/B} even better by applying much  stronger $\hat s$
cut. For example $\sqrt{\hat s}>300$ GeV in the case of  {\bf C2} set
suppresses the background by a factor of 8.7 while the QCD  signal only by a
factor of 2.6.

\begin{center}
\begin{tabular}{c|c|c|c}
\hline
$M_H=120$ GeV  & no $\sqrt{\hat s}$ cut & $\sqrt{\hat s}>210$ GeV
    & \raisebox{0ex}[3ex][1ex]{$\sqrt{\hat s}>300$ GeV}   \\ 
\hline
\raisebox{0ex}[3ex][2ex]{$\sigma_S^{QCD}$} fb &  10.6  & 7.0  & 4.0  \\
\raisebox{0ex}[0ex][2ex]{$\sigma_S^{EW}$}  fb &   1.1  & 0.81 & 0.58 \\
\raisebox{0ex}[1ex][1ex]{$\sigma_B$} fb/GeV   &   25.3 & 9.1  & 2.9  \\ 
\hline
\end{tabular}
\end{center}

The reducible QCD background should be also suppressed by the $\sqrt{\hat s}$
cut, even stronger than the irreducible one. Indeed, the subprocesses of
reducible background have a similar diagram topology. So our arguments (see
next section) should work in this case also. However, a shift to smaller
$\sqrt{\hat s}$ for the reducible background processes will be larger than for
irreducible ones, because the energy of parton misidentified as a photon will
be reconstructed with a much higher loss of energy in comparison with the photon
radiated from the hard subprocess directly.

\vskip 0.5cm
\noindent
{\large \bf 7. Angular distributions in the partonic c.m.s.}
\vskip 0.2cm

\noindent
Useful information is provided by the spin structure of {\it in} and {\it
out}-states. For the dominant subprocess $gg\to H+g$ a set of possible {\it in}
spin states does not include spin 1, while the spin of the {\it out-} state is
determined by the gluon. It means, in particular, that the S-wave does not
contribute here. At the same time for the background subprocesses
$gq\to\gamma+\gamma+q$ and $q\bar q\to\gamma+\gamma+g$  the same spin
configurations are possible for both {\it in} and {\it out} states. So, the jet
angular distribution in the partonic  c.m.s. should be different for the signal
and the background.

In Fig.~\ref{fig:aaj-thjstar}a we represent the angular distributions for the
signal and background processes. Here $\vartheta^*_{jet}$ is the jet scattering
angle in the partonic c.m.s. The background curve has a bump while the signal
curve is smooth enough. In Figs.~\ref{fig:aaj-thjstar}b and
\ref{fig:aaj-thjstar}c the same distributions are shown when the $\sqrt{\hat
s}$ cut is applied. One can see that events from the central part are
suppressed. Qualitative interpretation of this effect is given by the simple
observation that the relative contribution of partial waves with higher angular
momentum increases with the collision energy for processes with t-channel
exchange of virtual particles. Thus, the contribution of S-wave in the
background is relatively suppressed, substantiating the effect of the cut on
partonic collision energy $\sqrt{\hat s}$ discussed in the previous section.

It is clear that the angular distributions of photons from the signal and
background processes in the partonic c.m.s. should be different. Indeed,
photons from the Higgs decay (giving uniform angular distribution in the Higgs
rest frame) are produced mostly in the direction opposite to the jet in the
partonic c.m.s. At the same time photons radiated from the final quark in $gq$
subprocesses, which give dominant contribution to the background, will be
observed mainly at small angles with the jet. This effect is clearly seen in 
Fig.~\ref{fig:aaj-ja}a, where the distributions in $\vartheta^*_{g\gamma}$
(angle between  the jet and the photon with smaller $p_t$) are represented in
partonic c.m.s. Let us now apply the $\sqrt{\hat s}$ cut. The distributions in
this case are represented in  Figs.~\ref{fig:aaj-ja}b and \ref{fig:aaj-ja}c.
One can see that events at small angles $\vartheta^*_{g\gamma}$ are suppressed
affecting mainly the background distribution. This is qualitatively understood
because events with small angle $\vartheta^*_{g\gamma}$ correspond in average
to smaller momentum of the Higgs boson or $\gamma\gamma$ system. Indeed, for
high enough velocity of the $\gamma\gamma$ system the corresponding Lorentz
boost will turn (almost) all photons in the directions opposite to the jet
momentum. So, events where one photon is radiated in the jet hemisphere should
have smaller $\sqrt{\hat s}$ than events where both photons are radiated
opposite to the jet. In particular, this effect gives us one more argument why
the $\sqrt{\hat s}$ cut suppresses the background stronger than the signal.

\vskip 0.5cm
\noindent
{\large \bf 8. Estimates for the LHC detectors}
\vskip 0.2cm

\noindent
In this section we give some estimates of the {\it Signal/Background} ratio and
the signal significance for LHC detectors in the channel
$pp\to\gamma\gamma+jet$ using basic numbers from ATLAS and CMS Technical
Design Reports \cite{atlasTDR,cmsTDR}. First, the efficiency of photon
detection should be taken into account. This important instrumental parameter
is expected to be $\sim 80$\% for both detectors, and it was obtained from the
simulation of $pp\to H\to\gamma\gamma$ events in the detectors. Second, the
fiducial area cuts should be considered to exclude the regions of
electromagnetic calorimeter where the performance decreases crucially. The
corresponding efficiency per photon can be taken at the level 95\% for ATLAS
and 92.5\% for CMS. Finally, let us consider the $M_{\gamma\gamma}$ bin
optimization which affects rather strongly the signal significance. At LHC low
luminosity regime and for $M_H=100$ GeV the 80\% signal events reconstruction
efficiency corresponds to the $M_{\gamma \gamma}$ mass bin of 3.1 GeV 
(resolution parameter $\sigma_m=1.1$ GeV) for ATLAS Lead-Liquid-Argon
electromagnetic calorimeter \cite{atlasTDR}. For CMS PbWO$_4$ electromagnetic
calorimeter \cite{cmsTDR} the mass resolution parameter $\sigma_m=0.65$ GeV
assumes the mass bin $\Delta M_{\gamma\gamma}=1.9$ GeV with the reconstruction 
efficiency 73\% in this window. For larger $M_H$ the $M_{\gamma\gamma}$
resolution is slightly worse. Let us take the values $\Delta
M_{\gamma\gamma}=3.25$ GeV and $2.0$ GeV if $M_H=120$ GeV, and $\Delta
M_{\gamma\gamma}=3.4$ GeV and $2.1$ GeV if $M_H=140$ GeV, for  ATLAS and CMS
$M_{\gamma\gamma}$ bin, correspondingly.

Then, the QCD next-to-leading corrections should be taken into account.  As we
have mentioned above, these corrections are unknown for the processes discussed
here. Some estimate may be possible since it was shown (see \cite{SDGZ} and the
references therein) that the NLO corrections to the $gg$ fusion subprocesses in
the $pp\to\gamma\gamma+X$ inclusive channel increase the Higgs production cross
section by a factor $K \sim$ 1.6. One can hope that the corresponding
$K$-factor in our case will be at least of the same value. Of course, it is
very probable that the NLO corrections enhance somehow the rates of background
processes as well. So, in the absence of theoretical results let us use  in our
case the factor $K^{NLO}=1.6$ both for the signal and background subprocesses.

Finally we get the following numbers in the case of {\bf C2} basic kinematical
cuts and applying additional cut $\sqrt{\hat s}>300$ GeV:

\begin{center} 
\begin{tabular}{l|cc|cc|cc}
\hline
\raisebox{0ex}[2ex][1ex]{ Low} 
   & \multicolumn{2}{c|}{$M_H=100$ GeV \hspace{5mm}}
   & \multicolumn{2}{c|}{$M_H=120$ GeV \hspace{5mm}}
   & \multicolumn{2}{c}{$M_H=140$ GeV }\\ 
luminosity & ATLAS & CMS \hspace{5mm} & ATLAS & CMS \hspace{5mm} & ATLAS&CMS\\ 
\hline
\raisebox{0ex}[3ex][2ex]{$\sigma_S^{eff}$} \hspace{0.3cm} fb 
                                     & 1.8 & 1.6 & 3.4 & 2.9 & 3.6 & 3.1 \\ 
\raisebox{0ex}[0ex][2ex]{$\sigma_B^{eff}$} \hspace{0.3cm} fb 
                                     & 6.6 & 3.8 & 8.9 & 5.2 & 12.3& 7.2\\ 
\raisebox{0ex}[0ex][2ex]{$S/B$}         
                         & 1/3.7 & 1/2.5 & 1/2.7 & 1/1.8 & 1/3.4 & 1/2.3 \\ 
\raisebox{0ex}[1ex][1ex]{$S/\sqrt{B}$} (30 fb$^{-1}$)    
                                     & 3.8 & 4.3 & 6.2 & 7.0 & 5.6 & 6.3 \\
\hline 
\end{tabular} 
\end{center}

\vskip 0.5cm
\noindent
{\large \bf 9. Conclusions}
\vskip 0.2cm

\noindent  
The channel $\gamma\gamma+jet$ with the jet transverse energy $E_t>30$GeV and
rapidity $|\eta|<4.5$ (thus involving forward hadron calorimeters) gives very
promising discovery possibilities for the Higgs boson with a mass of 100-140
GeV during the LHC operation at a low luminosity of $\sim
10^{33}$cm$^{-2}$s$^{-1}$. For example, with an integrated luminosity of 30
fb$^{-1}$ about 100 signal events could be observed for $M_H=120$ GeV with a
number of background events only 3 times higher in ATLAS and 2 times higher in
CMS detector. These numbers demonstrate the main advantage of this channel in
comparison with the inclusive reaction $pp\to\gamma\gamma+X$, namely
significant improvement of the {\it S/B} ratio. The estimate of the reducible
background using only isolation criteria shows that it  is less than 20\% of
the irreducible background. Our results for the signal and background rates
mean that the discovery level $S/\sqrt{B}=5$ for the signal significance will
be achieved already with an integrated luminosity of 30 fb$^{-1}$ for
$M_H=110-140$ GeV.

We found that the detection of jets with rapidities up to $|\eta|=4.5$ improves
noticeably the signal significance. For example, 20-25\% improvement can be
achieved in comparison with the case when hard jets are only centrally
produced  ($|\eta|<2.4$, $E_t^{jet}>40$ GeV).

We demonstrated that sufficiently rich kinematics of the three particle final 
state allows to introduce new observable distributions, suitable for a better
separation of the signal. Further improvement of the signal significance can be
achieved by using the {\it jet} and  {\it jet-photon} angular distributions in
the reconstructed partonic c.m.s. These distributions can be also used for a
further suppression of the reducible QCD background.

One should also note that in the present analysis we used parameters obtained
from the simulation of the reaction $pp\to\gamma\gamma+X$ in ATLAS and CMS
detectors. However, the $\gamma\gamma+jet$ kinematics is more preferable for
the event reconstruction than in the inclusive case. Reconstruction of the jet
in the hadronic calorimeter allows to determine more precisely the position of
interaction vertex. Photons from the Higgs decay in $\gamma\gamma+jet$ state
are more energetic than for the inclusive channel. So the photon reconstruction
efficiency and effective mass resolution used in our analysis probably are too
pessimistic. From the experimental data processing point of view, additional
event selection criterion (trigger condition) of a jet in the final state
allows to restrict the number of diphoton events in comparison with the
inclusive channel, providing opportunities for a more careful analysis at a
better $S/B$ ratio. Alltogether these factors could give a sizeable improvement
of the signal significance.

\vskip 0.5cm
\noindent
{\large \bf Acknowledgements}
\vskip 0.2cm

We are grateful to E.~Boos, D.~Denegri, I.~Goloutvine, M.~Mangano,
A.~Nikitenko, A.~Pukhov and A.~Rozanov for many useful discussions. 

This work was partially supported by the Russian  Ministry of Science and
Technologies, INTAS grant 97-2085 and the Russian Foundation for Basic
Research (grants 98-02-17699 and 96-02-19773).

\newpage

\clearpage
{\large \bf Figure captions}
\vskip 0.5cm

\begin{figure}[hb]
\caption{Feynman diagrams for the QCD signal subprocesses:
a) $gg\to H+g$ , b) $gq\to H+q$ and c) $q\bar q\to H+g$.
\label{fig:fd_s_QCD} }

\caption{Feynman diagrams for the electroweak 
signal subprocesses with a) $WW$ and $ZZ$
fusion mechanisms of the Higgs boson production, b) associated $HW$
and $HZ$ production. 
\label{fig:fd_s_WZ} }

\caption{Feynman diagrams for the subprocesses
a) $gq\to\gamma+\gamma+q$ and b) $q\bar q\to\gamma+\gamma+g$
contributing to the irreducible background.
\label{fig:fd_Birr} }


\caption{Distributions in the parton c.m. energy ${\protect \sqrt{\hat s}}$
for the signal (S) (without EW contribution) and background (B) processes.  Here
$M_H=120$ GeV and the basic set of kinematical cuts {\bf C2} is imposed. The
$\gamma\gamma$ invariant mass for the background is integrated over the  1 GeV
bin.
\label{fig:aaj-shat}}

\caption{Distributions in the jet angle in partonic c.m.s.
for the signal (S) (without EW contribution) and background (B) processes  in 
the
case $M_H=120$ GeV. Upper plot is obtained with the basic  set of kinematical
cuts {\bf C2}. Next plots show changes in this distribution when the ${\protect
\sqrt{\hat s}}$ cut is applied. The $\gamma\gamma$ invariant mass for the
background is integrated over the  1 GeV bin.
\label{fig:aaj-thjstar}}

\caption{Distributions in the angle between jet and the photon with smaller
transverse momentum in partonic c.m.s. for signal (S) (without EW
contribution) and background (B) processes  in the case $M_H=120$ GeV. Upper
plot is obtained with the basic  set of kinematical cuts {\bf C2}. Next plots
show changes in this distribution when the ${\protect \sqrt{\hat s}}$ cut is
applied. The $\gamma\gamma$ invariant mass for the background is integrated
over the 1 GeV bin.
\label{fig:aaj-ja}}

\end{figure}

\clearpage
{\large \bf Figures}
\setcounter{figure}{0}

\vskip 2cm

\begin{figure}[hb]
\begin{center}
\unitlength=1cm
\vspace*{1cm}
\begin{picture}(16,8)
\put(0,9.5){\mbox{a)}}
\put(-1,4){\epsfxsize=22cm \leavevmode \epsfbox{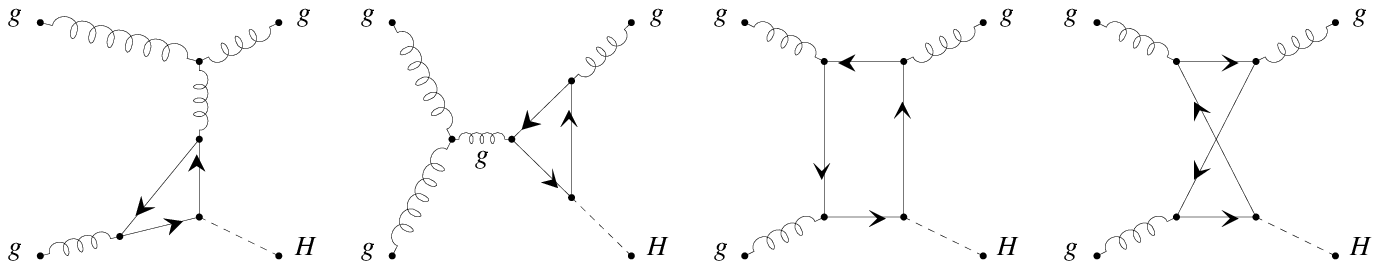}} 
\put(0,5){\mbox{b)}}
\put(0,0){\epsfxsize=22cm \leavevmode \epsfbox{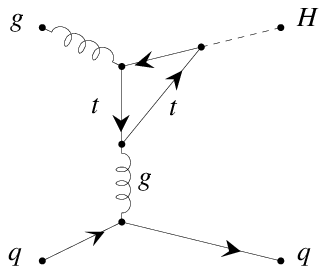}}
\put(8,5){\mbox{c)}}
\put(7.5,0){\epsfxsize=22cm \leavevmode \epsfbox{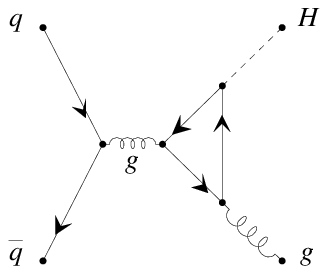}}
\end{picture}
\end{center}
\vspace*{-2cm}
\caption{}
\end{figure}

\begin{figure}[hb]
\begin{center}
\unitlength=1cm
\begin{picture}(16,8)
\put(-0.5,5.7){\mbox{a)}}
\put(-2,0){\epsfxsize=22cm \leavevmode \epsfbox{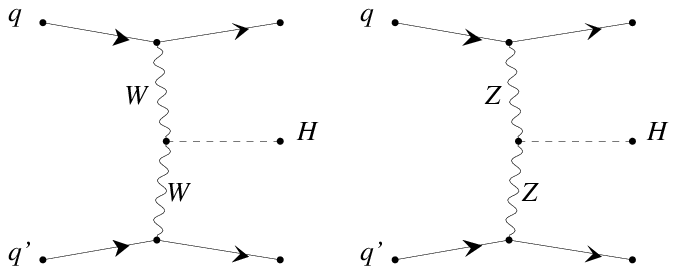}}
\put(7.5,5.7){\mbox{b)}}
\put(6.5,0){\epsfxsize=22cm \leavevmode \epsfbox{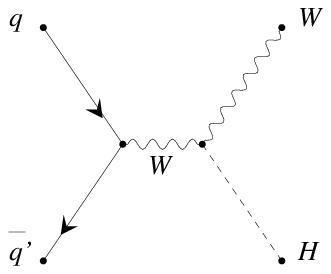}}
\put(10.5,0){\epsfxsize=22cm \leavevmode \epsfbox{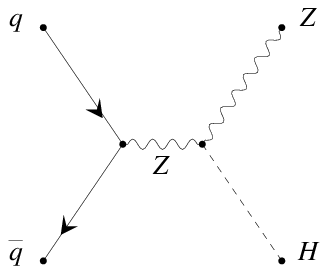}}
\end{picture}
\end{center}
\vspace*{-2cm}
\caption{}
\end{figure}

\clearpage

\begin{figure}[hb]
\unitlength=1cm
\begin{center}
\begin{picture}(16,8)
\put(1,9.5){\mbox{a)}}
\put(1,4){\epsfxsize=20cm \leavevmode \epsfbox{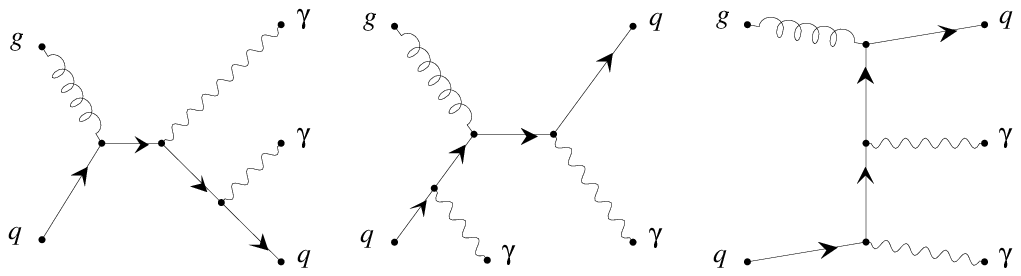}} 
\put(1,5){\mbox{b)}}
\put(1,0){\epsfxsize=20cm \leavevmode \epsfbox{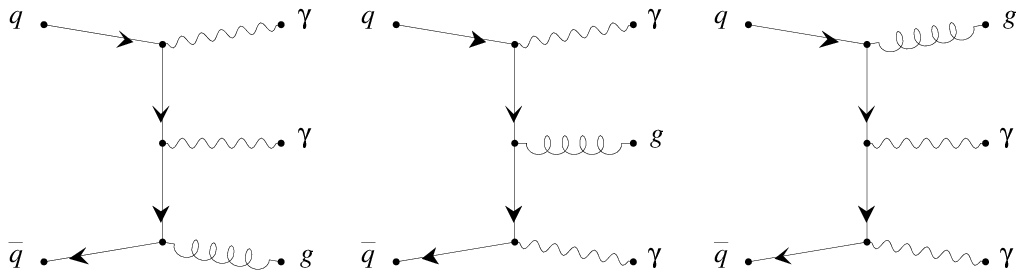}}
\end{picture}
\end{center}
\vspace*{-2cm}
\caption{}
\end{figure}

\begin{figure}[hb]
\begin{center}
\unitlength=1cm
\vspace*{-0.5cm}
\begin{picture}(16,11)
\put(2.5,10){\mbox{{\large $\protect \frac{d\sigma}{d\sqrt{\hat s}}$}
   \hspace{0.2cm}  fb/GeV}} 
\put(3,0){\epsfxsize=11cm \leavevmode \epsfbox{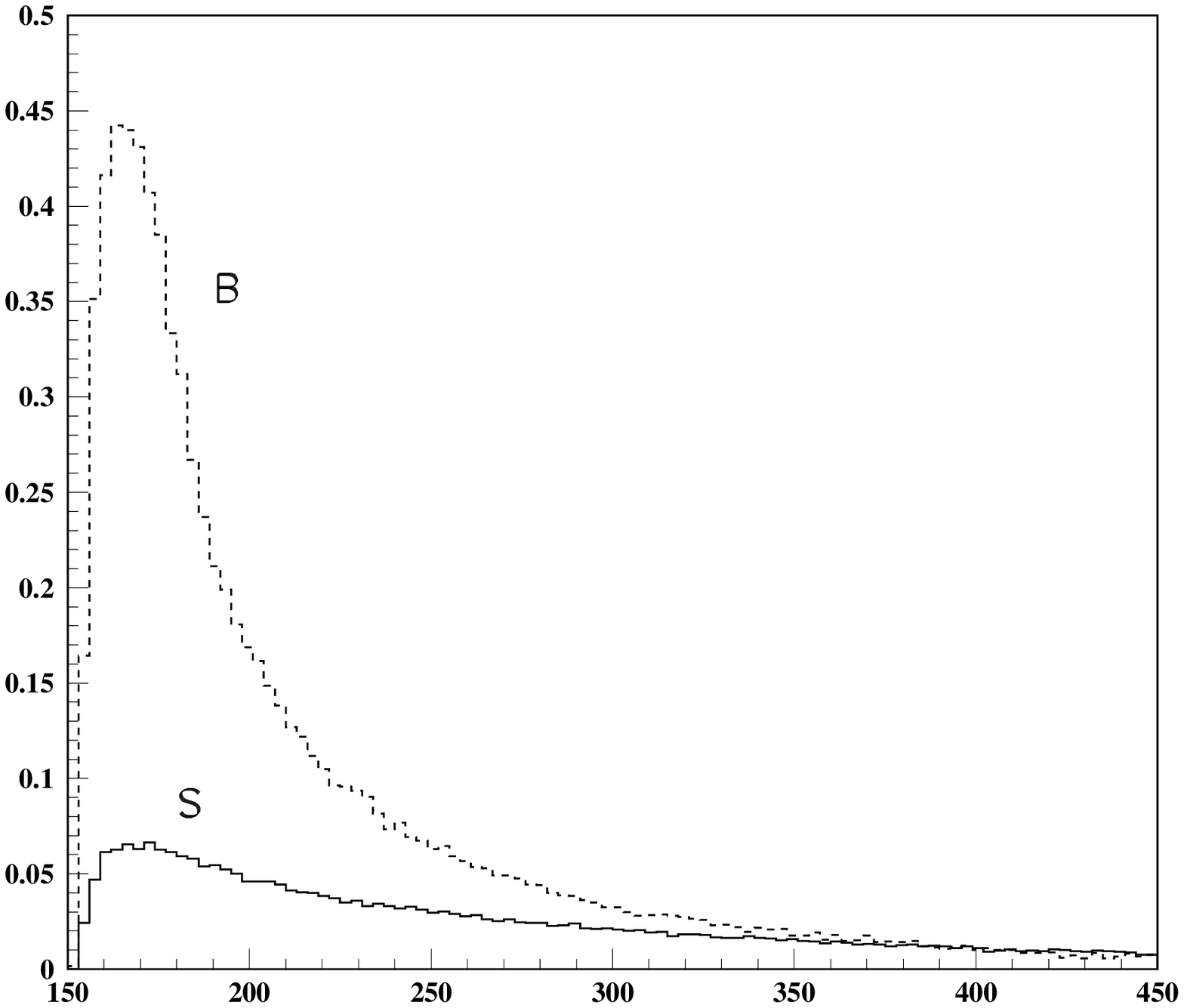}}
\put(11,0.5){\mbox{{ $\protect \sqrt{\hat s}$} \hspace{0.2cm} GeV}}
\end{picture}
\end{center}
\vspace*{-1cm}
\caption{}
\end{figure}

\clearpage

\begin{figure}[hb]
\begin{center}
\unitlength=1cm
\vspace*{2cm}
\begin{picture}(16,16)
\put(3.5,17){\mbox{{\LARGE $\protect \frac{d\sigma}{d\vartheta^*_{jet}}$} 
   \hspace{0.2cm} \large fb/deg}}
\put(2,16){\mbox{a)}}
\put(2,11){\mbox{b)}}
\put(2,5){\mbox{c)}}
\put(-2,0){\epsfxsize=16cm \leavevmode \epsfbox{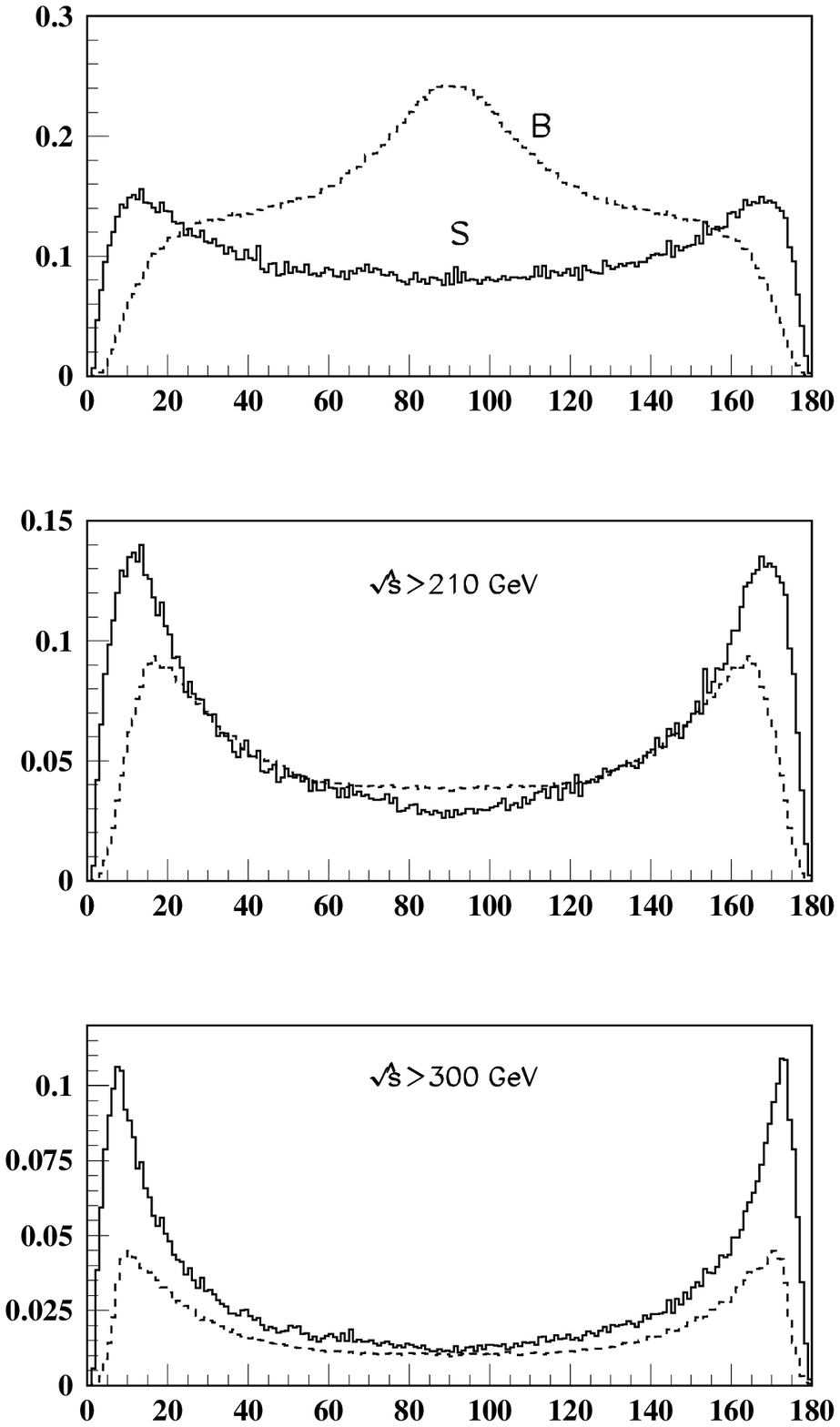}}
\put(11,-0.5){\mbox{{\Large $\protect \vartheta^*_{jet}$} \hspace{0.2cm} deg}}
\end{picture}
\end{center}
\vspace*{1cm}
\caption{}
\end{figure}

\clearpage

\begin{figure}[hb]
\begin{center}
\unitlength=1cm
\vspace*{2cm}
\begin{picture}(16,16)
\put(3.5,17){\mbox{{\LARGE $\protect \frac{d\sigma}{d\vartheta^*_{j\gamma}}$}
   \hspace{0.2cm} \large  fb/deg}}
\put(2,16){\mbox{a)}}
\put(2,10){\mbox{b)}}
\put(2,5){\mbox{c)}}
\put(-2,0){\epsfxsize=16cm \leavevmode \epsfbox{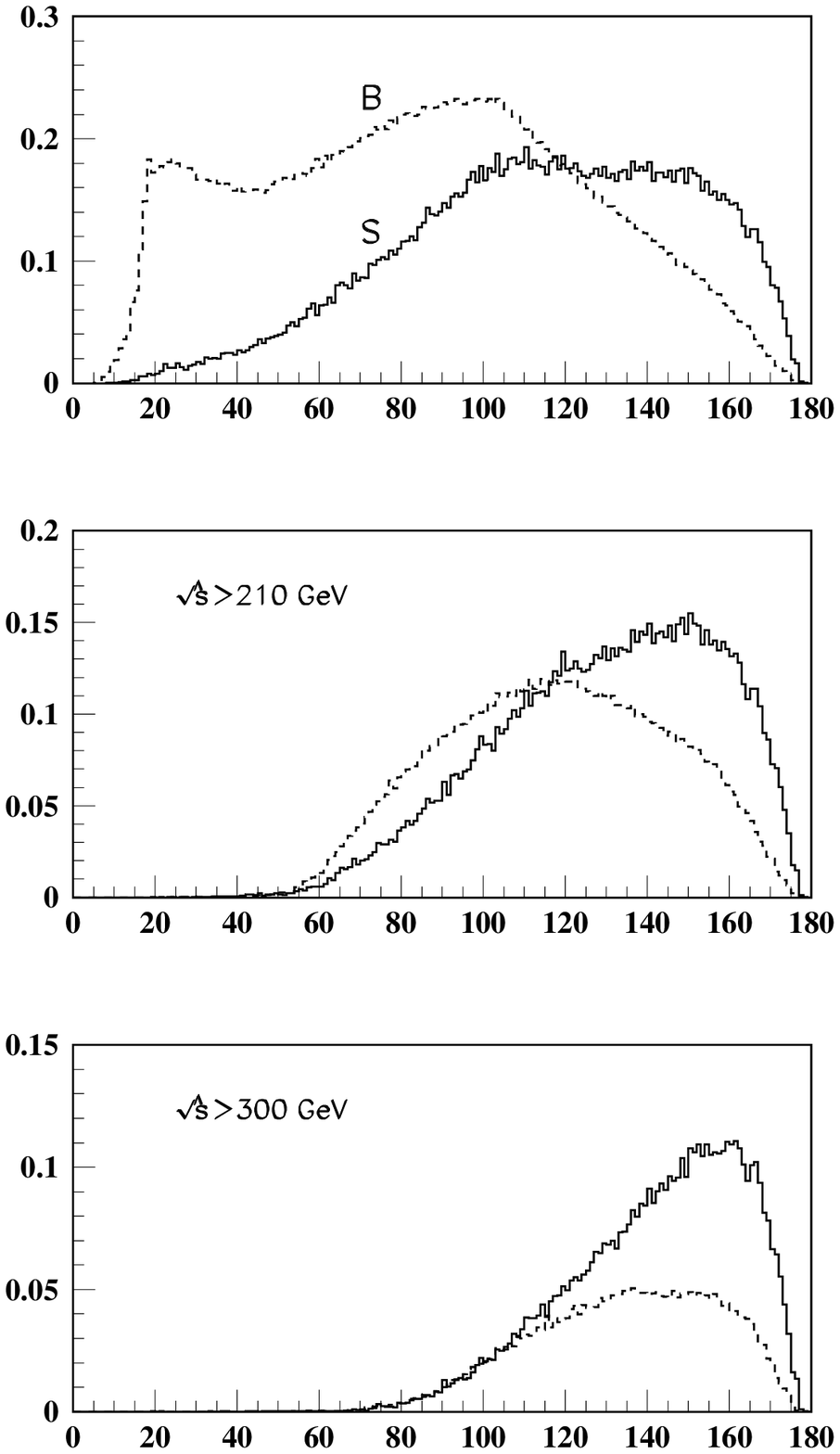}}
\put(11,-0.5){\mbox{{\large $\protect \vartheta^*_{j\gamma}$} \hspace{0.2cm} 
deg}}
\end{picture}
\end{center}
\vspace*{1cm}
\caption{}
\end{figure}

\end{document}